\documentclass[conference]{IEEEtran}
\usepackage{amsmath}
\usepackage{amssymb}
\usepackage{amsfonts}
\usepackage{amsthm}
\usepackage{algorithm}
\usepackage{algorithmic}
\usepackage{graphicx}
\usepackage{array}
\usepackage{lettrine}

\usepackage[shortlabels]{enumitem}
\usepackage{cite}
\usepackage{color}
\usepackage{epsfig,epstopdf}
\usepackage{url}
\usepackage{verbatim}
\usepackage{diagbox}
\usepackage{subfigure}
\usepackage{balance}
\usepackage{multirow}

\IEEEoverridecommandlockouts

\begin{document}

\title{Unsupervised Learning Based Hybrid Beamforming with Low-Resolution Phase Shifters for MU-MIMO Systems}

\author{\IEEEauthorblockN{Chia-Ho Kuo{\textsuperscript{$1$}}, Hsin-Yuan Chang{\textsuperscript{$1$}}, Ronald Y. Chang{\textsuperscript{$2$}}, and Wei-Ho Chung{\textsuperscript{$1,2$}}}
\IEEEauthorblockA{
{\textsuperscript{$1$}}Department of Electrical Engineering, National Tsing Hua University, Taiwan \\
{\textsuperscript{$2$}}Research Center for Information Technology Innovation, Academia Sinica, Taiwan \\
}
\IEEEauthorblockA{Email: c-ho.kuo@outlook.com, hyuan.chang@outlook.com, rchang@citi.sinica.edu.tw, whchung@ee.nthu.edu.tw}
\thanks{This work was supported by the Visible Project at the Research Center for Information Technology Innovation, Academia Sinica, and the Ministry of Science and Technology, Taiwan, under Grants MOST 109-2221-E-001-013-MY3 and MOST 110-2221-E-007-042-MY3.}
}

\maketitle

\begin{abstract}
Millimeter wave (mmWave) is a key technology for fifth-generation (5G) and beyond communications. Hybrid beamforming has been proposed for large-scale antenna systems in mmWave communications. Existing hybrid beamforming designs based on infinite-resolution phase shifters (PSs) are impractical due to hardware cost and power consumption. In this paper, we propose an unsupervised-learning-based scheme to jointly design the analog precoder and combiner with low-resolution PSs for multiuser multiple-input multiple-output (MU-MIMO) systems. We transform the analog precoder and combiner design problem into a phase classification problem and propose a generic neural network architecture, termed the phase classification network (PCNet), capable of producing solutions of various PS resolutions. Simulation results demonstrate the superior sum-rate and complexity performance of the proposed scheme, as compared to state-of-the-art hybrid beamforming designs for the most commonly used low-resolution PS configurations.
\end{abstract}


\IEEEpeerreviewmaketitle

\section{Introduction}

Millimeter wave (mmWave) communication has been proposed as a promising technology to meet the requirements of high data rate and low latency in fifth-generation (5G) communications \cite{mmwave}. However, mmWave communication suffers from higher weather sensitivity, propagation loss, and penetration loss. Fortunately, the inter-antenna spacing is proportional to the wavelength of mmWave frequencies. Thus, the mmWave systems enable large-scale antenna arrays to provide significant beamforming gain to compensate these losses. Hybrid beamforming for mmWave communication systems with large-scale antenna arrays design has been recently studied \cite{Heq1,Limited,MMSE1}. The architecture of hybrid beamforming comprises low-dimensional digital beamforming and high-dimensional analog beamforming, where the latter is realized by phase shifters.

In \cite{Heq1}, a hybrid block diagonalization (BD) scheme was proposed for massive MIMO systems to harvest the large array gain in the analog precoder and combiner, where the interference was canceled by the baseband BD solution. In \cite{Limited}, a two-stage hybrid precoding was proposed, first obtaining the analog precoder and combiner, and then the zero-forcing (ZF) baseband precoder. In \cite{MMSE1}, a new hybrid precoding design was developed to minimize the mean-squared error (MSE) of all data stream via the orthogonal matching pursuit (OMP) algorithm. The aforementioned works assume that analog precoder and combiner are implemented by infinite-resolution phase shifters (PSs). However, it is impractical to implement infinite-resolution PSs due to high hardware cost and power consumption. A direct approach to designing the analog precoder and combiner with low-resolution PSs is to directly quantize the elements of the analog precoder and combiner obtained under the condition of infinite resolutions \cite{quantizeInf}. This approach is however not efficient for designing low-resolution PSs. 

%%%%%%%%%%%%%%%%%%%%%%%%%%%%%%%%%%%%%%%%%%%%%%%%%%%%%%%%%%%%%%%%%
\begin{figure*}[t]
\centering
\includegraphics[width=0.7\textwidth]{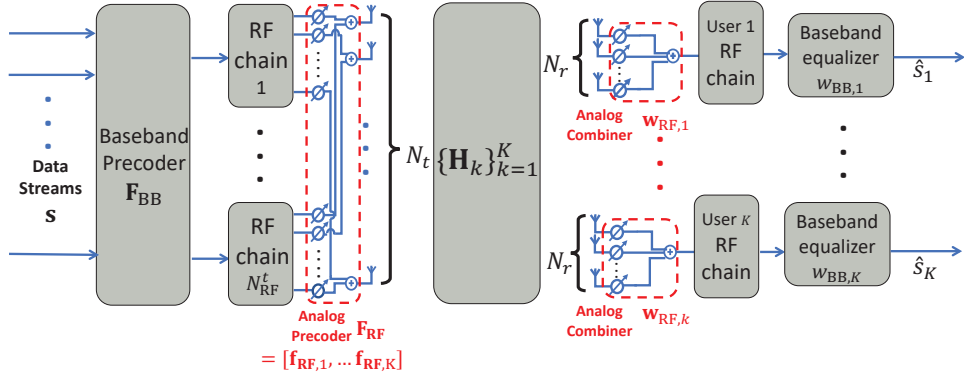}
\caption{System diagram of a downlink multiuser mmWave massive MIMO system with hybrid beamforming.}
\label{fig:exp}
\end{figure*}
%%%%%%%%%%%%%%%%%%%%%%%%%%%%%%%%%%%%%%%%%%%%%%%%%%%%%%%%%%%%%%%%%

Alternative methods \cite{CEO,DS_MISO,low} have been proposed to design the analog  precoder and combiner with low-resolution PSs. In \cite{CEO}, a cross-entropy-based algorithm with low-resolution PS was proposed for single-user and multiuser scenarios. In \cite{DS_MISO}, a hybrid beamforming scheme with dynamic subarrays and low-resolution PSs for multiuser multiple-input single-output (MU-MISO) was proposed, where dynamic connecting was introduced to mitigate the performance loss due to the use of low-resolution PS. In \cite{low}, a low-complexity hybrid precoding and combining design with two-bit resolution PSs for MU-MIMO systems was developed. However, the aforementioned optimization-based algorithms face different challenges in efficiently constructing the hybrid beamforming matrices. 

In this paper, we introduce machine learning based approach to designing hybrid beamforming with low-resolution PSs for MU-MIMO systems. We propose a deep learning approach to designing the analog precoder/combiner, coupled with a minimum mean-squared error (MMSE)-based baseband precoder. A concatenated neural network structure is proposed to facilitate arbitrary resolution PSs design and achieve better performance. Numerical results show that the proposed algorithm achieves favorable sum-rate performance for two-bit and three-bit PSs with low computational complexity, as compared to other hybrid beamforming algorithms.

The remainder of the paper is organized as follows. In Section~\ref{sec:system}, the signal model and the considered problem are described in detail. Section~\ref{sec:method} introduces the proposed unsupervised learning scheme for beamforming. The simulation results are presented in Section~\ref{sec:results} as well as in-depth discussions, followed by concluding remarks in Section~\ref{sec:conclusion}.

{\it Notations}: ${\mathbf M}(m,n)$ and ${\mathbf M}(m,:)$ denote the elements in the $m$th row and $n$th column, and the $m$th row of matrix ${\mathbf M}$, respectively. $\|\cdot\|_0$ and $\|\cdot\|_2$ denote the $l_0$-norm and $l_2$-norm of a vector. $(\cdot)^{T}$, $(\cdot)^{H}$, $(\cdot)^{-1}$, and $\|\cdot\|_{F}$ denote the transpose, conjugate-transpose, inverse, and Frobenius norm of a matrix, respectively. $|\cdot|$ denotes the absolute value.

\section{System Model} 
\label{sec:system}

\subsection{Signal Model}

We consider a downlink multiuser mmWave massive MIMO system employing hybrid beamforming, as illustrated in Fig.~\ref{fig:exp}. A base station (BS) with $N_t$ antennas and $N_t^{\rm RF}$ RF chains serves $K$ mobile stations (MSs), where each MS is equipped with $N_r$ antennas and $N_r^{\rm RF}$ RF chains. In this paper, to reduce hardware cost and decrease power consumption, we consider that each MS only supports one data stream, i.e., $N^{\rm RF}_r=1$. In the hybrid beamforming structure, the $K\times1$ transmitted symbols ${\mathbf s}=[s_1,s_2,\ldots,s_K]^{\it{T}}$ where $\Bbb{E} \{ {\mathbf s} {\mathbf s}^{\it{H}} \} = \frac{P}{K}{\mathbf I}_{K}$ and $P$ is the average total transmitted power, are precoded using a $N_t^{\rm RF} \times K$ baseband precoder ${\mathbf F}_{\rm BB}=[ {\mathbf f}_{{\rm BB}_1},{\mathbf f}_{{\rm BB}_2},\ldots,{\mathbf f}_{{\rm BB}_K} ]$ where ${\mathbf f}_{{\rm BB}_k}$ is the baseband precoder vector for the $k$th transmitted symbol, as well as an $N_t \times N_t^{\rm RF}$ analog precoder ${\mathbf F}_{\rm RF}=[ {\mathbf f}_{{\rm RF}_1},{\mathbf f}_{{\rm RF}_2},\ldots,{\mathbf f}_{{\rm RF}_K}]$. The transmitted signal can be written as $\mathbf{x} = {\mathbf F}_{\rm RF} {\mathbf F}_{\rm BB}{\mathbf s}$, with the normalized power constraint $\|{\mathbf F}_{\rm RF}{\mathbf F}_{\rm BB}\|^2_{\it{F}} = K$. We adopt a narrowband block-fading channel model and the signal received by the $k$th MS can be expressed as 
    \begin{equation}
    {\mathbf y}_k = {\mathbf H}_k\mathbf{x}+\mathbf{n}_k
    \end{equation}
where $\mathbf{H}_k$ is an $N_r\times N_t$ matrix that represents the downlink channel from the BS to the $k$th MS, and $\mathbf{n}_k$ is an additive complex Gaussian white noise with zero mean and covariance matrix $\sigma^{2}\mathbf{I}$, i.e., $\mathbf{n}_k \sim \mathcal{CN} (0,\,\sigma^{2}\mathbf{I})$. Then, an $N_r \times 1$ analog combiner ${\mathbf w}_{{\rm RF}_k}$ and a baseband equalizer $w_{{\rm BB}_k}$ are used to process the received signal $\mathbf{y}_k$. The digital combined signal at the $k$th MS is given by
    \begin{equation}
    \begin{split} 
    &\widehat{\mathbf{s}}_k =  ({\mathbf w}_{{\rm RF}_k} w_{{\rm BB}_k})^{\it{H}} {\mathbf H}_k {\mathbf F}_{\rm RF} {\mathbf f}_{{\rm BB}_k}{\mathbf s}_k \\
    &+\sum_{j\neq k}^{K}  ({\mathbf w}_{{\rm RF}_k} w_{{\rm BB}_k})^{H} {\mathbf H}_k {\mathbf F}_{\rm RF} {\mathbf f}_{{\rm BB}_j} {\mathbf s}_j
    +({\mathbf w}_{{\rm RF}_k} w_{{\rm BB}_k})^{\it{H}} {\mathbf n}_k.
    \end{split}
    \end{equation}
The signal-to-interference-plus-noise ratio (SINR) of the $k$th MS can be expressed as
    \begin{equation} \label{eq:SINR}
    \begin{split} 
    &{\rm SINR}_k =\\ 
    &  \frac{\frac{P}{K}|({\mathbf w}_{{\rm RF}_k}w_{{\rm BB}_k})^{\it{H}} \mathbf{H}_k {\mathbf F}_{\rm RF} {\mathbf f}_{{\rm BB}_k}|^2} {\frac{P}{K}\sum_{j\neq k}^{K}|({\mathbf w}_{{\rm RF}_k}w_{{\rm BB}_k})^{\it{H}} \mathbf{H}_k {\mathbf F}_{\rm RF} {\mathbf f}_{{\rm BB}_j}|^2 + \sigma^{2}\|{\mathbf w}_{{\rm RF}_k}w_{{\rm BB}_k}\|^2_2}.
    \end{split}
    \end{equation}
The achievable sum-rate of the system is $R_{\rm{sum}} = \sum_{k=1}^{K}\log_{2}(1+{\rm SINR}_k)$.

We consider that the analog precoder ${\mathbf F}_{\rm RF}$ and combiner ${\mathbf w}_{{\rm RF}_k}$ are implemented with finite-resolution analog phase-shifters (PSs). Specifically, the elements of ${\mathbf F}_{\rm RF}$ and ${\mathbf w}_{{\rm RF}_k}$ are subject to a constant modulus constraint with phases being restricted to a $B$-bit finite discrete phase set, i.e., ${\mathbf F}_{\rm RF}(m,n) \in \mathcal{F} \triangleq \{ \frac{1}{\sqrt{N_t}} \mathrm{e}^{j\frac{2\pi b}{2^B}}|b=0,1,\ldots,2^B-1\}$, for $m=1,\ldots,N_t^{}$ and $n=1,\ldots,N_t^{\rm RF}$, and ${\mathbf w}_{{\rm RF}_k}(n) \in \mathcal{W} \triangleq \{\frac{1}{\sqrt{N_r}}\mathrm{e}^{j\frac{2\pi b}{2^B}}|b=0,1,\ldots,2^B-1\}$, for $n=1,\ldots,N_r^{}$.

\subsection{mmWave Channel Model}

We model the mmWave channel by the extended Saleh-Valenzuela channel model \cite{MMSE1,MMSE2}. The channel between the BS and the $k$th MS can be expressed as
    \begin{equation}
    \mathbf {H}_k = 
    \sqrt{\frac{N_t N_r}{L_{k}}}
    \sum_{l=1}^{L_{k}} \alpha_{k,l}
    {\mathbf a_{r}}(\phi_{k,l}^{r},\theta_{k,l}^{r}) {\mathbf a}_{t}^{H}(\phi_{k,l}^{t},\theta_{k,l}^{t})
    \end{equation}
where $L_{k}$ is the number of propagation paths, $\alpha_{k,l}$ is the complex gain of the {$l$th} path between the BS and the $k$th MS. It is assumed that $\alpha_{k,l}$ are i.i.d. complex Gaussian ${\cal CN}(0,1)$. $\phi_{k,l}^{r}$ ($\theta_{k,l}^{r}$) and $\phi_{k,l}^{t}$ ($\theta_{k,l}^{t}$) represent the azimuth (elevation) angles of arrival and departure, respectively. For a uniform planar array (UPA) with $\sqrt{N} \times \sqrt{N}$ elements, the antenna array response vector can be written as
    \begin{align}
    &\mathbf{a}(\phi_{k,l},\theta_{k,l}) \nonumber\\
    &= \frac{1}{\sqrt{N}}\Big[1,\ldots,{\rm e}^{j\frac{2\pi}{\lambda}d(m\sin{(\phi_{k,l}})\sin{(\theta_{k,l})}+n\cos{(\theta_{k,l}))}}, \nonumber\\
    &\quad\ldots,{\rm e}^{j\frac{2\pi}{\lambda}d((\sqrt{N}-1)\sin{(\phi_{k,l})}\sin{(\theta_{k,l})}+(\sqrt{N}-1)\cos{(\theta_{k,l})})}\Big]^{T}, \nonumber\\
    &\qquad\qquad\qquad\qquad m,n=0,1,\ldots,\sqrt{N}-1
    \end{align}
where $\lambda$ is the wavelength, $d$ is the antenna spacing, and $m,n$ are the antenna indices of the 2D plane. In this paper, we assume that perfect channel state information (CSI) is available.

\subsection{Problem Formulation}

Our objective is to design the baseband precoder, analog precoder, baseband equalizer, and analog combiner to maximize the sum-rate of a multiuser mmWave system. Since the baseband equalizers $w_{{\rm BB}_k}$ in the numerator and denominator of \eqref{eq:SINR} cancel out, they have no effect on the sum-rate. This leads to the following design problem:  
\begin{subequations} \label{opt_1}
\begin{align}
     \underset{{\mathbf F}_{\rm RF},{\mathbf F}_{\rm BB},{\mathbf w}_{{\rm RF}_k}}{\arg\max}
    & \ \ R_{\rm{sum}} \\
     \text{s.t.} \ \  
    & \ \  {\mathbf F}_{\rm RF}(m,n) \in \mathcal{F} \label{opt_con_1} \\
    & \ \  {\mathbf w}_{{\rm RF}_k}(n) \in \mathcal{W} \label{opt_con_2} \\
    &\ \ \|{\mathbf F}_{\rm RF}{\mathbf F}_{\rm BB}\|^2_{\it{F}} = K. \label{opt_con_3}
\end{align}
\end{subequations}
Problem~\eqref{opt_1} is non-convex due to constraints \eqref{opt_con_1}--\eqref{opt_con_3}. Since \eqref{opt_con_1} and \eqref{opt_con_2} constrain to a finite discrete set, theoretically, the optimal solutions can be found by an exhaustive search. However, the set of candidate solutions grows exponentially with the numbers of antennas and RF chains at the transmitter, the number of MSs, and the resolution of analog beamforming, i.e., $2^{B\times N_t \times N_t^{\rm RF}\times B\times N_r\times K}$. Feasible methods have been proposed to design the analog precoder and combiner with fixed-resolution PSs  \cite{low,quantizeInf}. The method proposed in \cite{low} is based on successively designing beamforming for multiple users and applies to only 2-bit PSs. The method proposed in \cite{quantizeInf} is based on directly quantizing the elements of the optimal analog precoder and combiner obtained under the condition of infinite-resolution PSs. This method, while applicable to arbitrary resolution PSs, incurs some performance degradation. To address these issues, we propose a DL-based method that can be implemented with arbitrary PS resolutions while achieving satisfactory performance.

%%%%%%%%%%%%%%%%%%%%%%%%%%%%%%%%%%%%%%%%%%%%%%%%%%%%%%%%%%%%%%%%%
\begin{figure*}[t]
\centering
\includegraphics[width=0.98\textwidth]{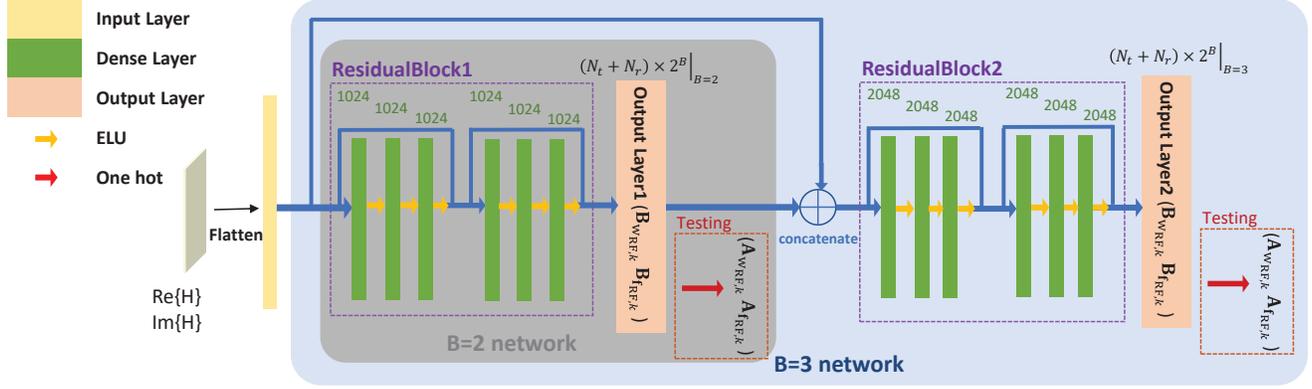}
\caption{The architecture of the proposed PCNet, illustrated for $B=2$ and $B=3$. The networks used to produce $B=2$ and $B=3$ solutions are shown in the gray and blue shaded areas, respectively.}
\label{fig:model}
\end{figure*}
%%%%%%%%%%%%%%%%%%%%%%%%%%%%%%%%%%%%%%%%%%%%%%%%%%%%%%%%%%%%%%%%%

\section{The Proposed Hybrid Beamforming Design}
\label{sec:method}
To solve the challenging problem~\eqref{opt_1}, we adopt a two-stage algorithm similar to \cite{Limited} where we divide the original problem into two subproblems and solve each subproblem in each stage. This method eases design difficulties and yields acceptable performance. In this section, we first introduce the two-stage algorithm and then the proposed DL-based realization of the algorithm. 

\subsection{The Two-Stage Algorithm}

The two-stage algorithm is based on successively designing the analog precoder and combiner for each user to maximize each user's signal power while neglecting the multiuser interference in the first stage, and designing the baseband precoder to address the multiuser interference in the second stage, as described as follows.

\subsubsection{The First Stage}

The analog precoder and combiner design problem in the first stage is described as
\begin{subequations} \label{opt_two_stage}
\begin{align}
     \underset{{\mathbf f}_{{\rm RF}_k},{\mathbf w}_{{\rm RF}_k}}{\arg\max}
    & \ \ \left|({\mathbf w}_{{\rm RF}_k})^{\it{H}} \mathbf{H}_k {\mathbf f}_{{\rm RF}_k} \right| \\
     \text{s.t.} \ \
    & \ \  {\mathbf f}_{{\rm RF}_k}(n) \in \mathcal{F} , \  {\mathbf w}_{{\rm RF}_k}(n) \in \mathcal{W}.  \label{opt_two_stage_con_1}
\end{align}
\end{subequations}
In \eqref{opt_two_stage}, we design ${\mathbf f}_{{\rm RF}_k}$ and ${\mathbf w}_{{\rm RF}_k}$ sequentially for the $k$th MS ($k=1,2,\dots,K$) by maximizing the signal power for each MS. The objective function of \eqref{opt_two_stage} derives directly from \eqref{eq:SINR} by neglecting the multiuser interference and noise in the denominator, and with a fixed baseband precoder in the numerator.

\subsubsection{The Second Stage}

After obtaining ${\mathbf f}_{{\rm RF}_k}$ and ${\mathbf w}_{{\rm RF}_k}$ for all users in the first stage, the baseband precoder ${\mathbf F}_{{\rm BB}}$ is designed to tackle the multiuser interference in the second stage. With fixed ${\mathbf f}_{{\rm RF}_k}$ and ${\mathbf w}_{{\rm RF}_k}$, we can consider the transmitter RF chain ${\mathbf F}_{\rm RF}$, the wireless channel ${\mathbf H}_k$, and the receiver RF chain ${\mathbf w}_{{\rm RF}_k}$ together as the equivalent channel for the $k$th MS \cite{Heq1,Heq2}, denoted by ${\mathbf h}_{{\rm eq}_{k}} =(({\mathbf w}_{{\rm RF}_k})^{\it{H}} \mathbf{H}_k {\mathbf F}_{\rm RF})^{\it H}$. Define ${\mathbf H}_{\rm eq}=[{\mathbf h}_{{\rm eq}_{1}},\ldots,{\mathbf h}_{{\rm eq}_{K}}]$. Then, the baseband precoder is designed based on the MMSE criterion, i.e.,
\begin{equation} 
{\mathbf F}_{{\rm BB}} = \left({\mathbf H}_{\rm eq}{\mathbf H}_{\rm eq}^{\it{H}}+\frac{K\sigma^2}{P}
{\mathbf F}_{\rm RF}{\mathbf F}_{\rm RF}^{\it{H}}
\right)^{-1}{\mathbf H}_{\rm eq}.
\end{equation}

\subsection{The Proposed Phase Classification Network (PCNet)-Based Analog Precoder and Combiner Design}
 
We propose a DL approach to solving the first-stage problem, i.e., problem~\eqref{opt_two_stage}. We first perform a problem reformulation. Let ${\mathbf p}_{{\mathbf f}_{{\rm RF}_k}} \triangleq \frac{1}{\sqrt{N_t}}[1,\ldots,\mathrm{e}^{j\frac{2\pi (2^B-1)}{2^B}}]^{\it{T}}$ and ${\mathbf p}_{{\mathbf w}_{{\rm RF}_k}} \triangleq \frac{1}{\sqrt{N_r}}[1,\ldots,\mathrm{e}^{j\frac{2\pi (2^B-1)}{2^B}}]^{\it{T}}$ be $2^B\times 1$ vectors containing all the elements in ${\cal F}$ and ${\cal W}$, respectively, and let ${\mathbf A}_{{\mathbf f}_{{\rm RF}_k}}$ and ${\mathbf A}_{{\mathbf w}_{{\rm RF}_k}}$ be $N_t\times 2^B$ and $N_r\times 2^B$ binary matrices, respectively. Then, by use of the relations ${\mathbf f}_{{\rm RF}_k}={\mathbf A}_{{\mathbf f}_{{\rm RF}_k}}{\mathbf p}_{{\mathbf f}_{{\rm RF}_k}}$ and ${\mathbf w}_{{\rm RF}_k}={\mathbf A}_{{\mathbf w}_{{\rm RF}_k}}{\mathbf p}_{{\mathbf w}_{{\rm RF}_k}}$, designing ${\mathbf f}_{{\rm RF}_k}$ and ${\mathbf w}_{{\rm RF}_k}$ in problem~\eqref{opt_two_stage} translates to designing ${\mathbf A}_{{\mathbf f}_{{\rm RF}_k}}$ and ${\mathbf A}_{{\mathbf w}_{{\rm RF}_k}}$ in the following equivalent problem:
\begin{subequations} \label{opt_2}
\begin{align}
     \underset{{\mathbf A}_{{\mathbf f}_{{\rm RF}_k}},{\mathbf A}_{{\mathbf w}_{{\rm RF}_k}}}{\arg\max}
    & \ \ \left|({\mathbf A}_{{\mathbf w}_{{\rm RF}_k}}{\mathbf p}_{{\mathbf w}_{{\rm RF}_k}})^{\it{H}} \mathbf{H}_k ({\mathbf A}_{{\mathbf f}_{{\rm RF}_k}}{\mathbf p}_{{\mathbf f}_{{\rm RF}_k}}) \right| \label{class_obj}\\
     \text{s.t.} \ \
    & \ \ {\mathbf A}_{{\mathbf f}_{{\rm RF}_k}}\in \{0,1\}^{N_t\times {2^B}}, \label{class_con_1}\\
    &\ \ {\mathbf A}_{{\mathbf w}_{{\rm RF}_k}} \in \{0,1\}^{N_r \times {2^B}}, \label{class_con_2} \\
    & \ \ \|{\mathbf A}_{{\mathbf f}_{{\rm RF}_k}}(m,:)\|_0=1,  \ \ \forall m, \label{class_con_3} \\ 
    & \ \ \|{\mathbf A}_{{\mathbf w}_{{\rm RF}_k}}(m,:)\|_0=1,  \ \ \forall m.    \label{class_con_4}
\end{align}
\end{subequations}
Constraints \eqref{class_con_3} and \eqref{class_con_4} state that each row of ${\mathbf A}_{{\mathbf f}_{{\rm RF}_k}}$ and ${\mathbf A}_{{\mathbf w}_{{\rm RF}_k}}$ has precisely one nonzero element, which, in combination with constraints \eqref{class_con_1} and \eqref{class_con_2}, establish \eqref{opt_two_stage_con_1}.

Designing the phases out of a finite discrete set for the analog precoder and combiner is similar to predicting discrete class labels (i.e., phases) in a classification problem. Thus, we propose to use a deep neural network (DNN) that performs phase classification to solve problem~\eqref{opt_2}. A deep learning (DL) approach presents several advantages in this specific problem. First, the complexity can be prohibitively high in solving the combinatorial optimization problem in problem~\eqref{opt_2} for large systems, and a DL approach could provide low-complexity, high-performance solutions to achieve real-time hybrid beamforming. Second, unlike some conventional methods that were developed for specific PS resolutions and not directly applicable to other PS resolutions, a DL approach could provide a generic framework applicable to various PS resolutions. 

The proposed unsupervised-learning-based phase classification network (PCNet) to solve problem~\eqref{opt_2} is depicted in Fig.~\ref{fig:model}. Since PSs with resolutions of two and three bits are most commonly and practically used \cite{PSs_1,PSs_2,PSs_3}, we consider $B=2$ and $B=3$ here for both illustrative and practical purposes. However, the proposed method can be readily extended to PSs with arbitrary resolutions. As shown in Fig.~\ref{fig:model}, for the case of $B=2$, the channel matrix $\mathbf {H}_k$ is first transformed to an equivalent real representation and fed into a residual block, termed ResidualBlock1, as inspired by the ResNet \cite{Residual}. The objective of ResidualBlock1 is to extract essential features that are helpful for NN to design PSs. Specifically, we adopt skip connections to fuse high-level hidden features acquired in the extraction process and low-level ones from early layers. Such a design aids the feature extraction in ResidualBlock1. ResidualBlock1 contains six fully-connected layers with $1024$ neurons each layer and Exponential Linear Unit (ELU) activation and dropout with dropout probability $0.3$. Two skip connections are exploited to sum the features. Then, the output of ResidualBlock1 is accepted as input by OutputLayer1, a fully-connected layer with $(N_t+N_r)\times2^B \big|_{B=2}$ neurons. In OutputLayer1, the neurons are resized into two matrices ${\mathbf B}_{{\mathbf f}_{{\rm RF}_k}}$ and ${\mathbf B}_{{\mathbf w}_{{\rm RF}_k}}$ of dimensions $N_t\times2^B\big|_{B=2}$ and $N_r\times2^B\big|_{B=2}$, respectively, followed by the $\rm{softmax}(\cdot)$ operation performed in a row-wise manner on ${\mathbf B}_{{\mathbf f}_{{\rm RF}_k}}$ and ${\mathbf B}_{{\mathbf w}_{{\rm RF}_k}}$. During the training process, the probability distributions over classes produced by $\rm{softmax}(\cdot)$ in ${\mathbf B}_{{\mathbf f}_{{\rm RF}_k}}$ and ${\mathbf B}_{{\mathbf w}_{{\rm RF}_k}}$ are employed to calculate the loss for backpropagation. After the training process is completed, during the testing, the probability distribution will be processed with a one-hot function to predict exactly one phase value for each PS. Specifically, ${\mathbf A}_{{\mathbf f}_{{\rm RF}_k}}(m,n)=1$ where $n={\arg\max}~{\rm softmax}({\mathbf B}_{{\mathbf f}_{{\rm RF}_k}}(m,:))$, and ${\mathbf A}_{{\mathbf f}_{{\rm RF}_k}}(m,n')=0, \forall n'\neq n$. Likewise, ${\mathbf A}_{{\mathbf w}_{{\rm RF}_k}}(m,n)=1$ where $n={\arg\max}~{\rm softmax}({\mathbf B}_{{\mathbf w}_{{\rm RF}_k}}(m,:))$, and ${\mathbf A}_{{\mathbf w}_{{\rm RF}_k}}(m,n')=0, \forall n'\neq n$. Then, the solutions to problem~\eqref{opt_2} for $B=2$ are given by the one-hot encoded output of OutputLayer1.

The proposed PCNet incorporates a concatenated architecture, where the solutions for $B=2$ are leveraged to produce solutions for $B=3$. As shown in Fig.~\ref{fig:model}, the channel matrix $\mathbf {H}_k$, as well as the results ${\mathbf B}_{{\mathbf f}_{{\rm RF}_k}}$ and ${\mathbf B}_{{\mathbf w}_{{\rm RF}_k}}$ produced by OutputLayer1 for the case of $B=2$, are fed into ResidualBlock2. ResidualBlock2 essentially has the same structure as ResidualBlock1 but more ($2048$) neurons per layer. Subsequent operations are similar to the case of $B=2$, and the one-hot encoded output of OutputLayer2 produces the solutions to problem~\eqref{opt_2} for $B=3$. The proposed concatenated architecture leverages the additional information provided by the lower-resolution solutions to the same problem, which leads to better higher-resolution solutions and faster convergence in network training as compared to training an independent network for each resolution independently. Furthermore, the concatenated architecture provides a general framework for designing PSs with arbitrary resolutions. Specifically, networks corresponding to lower resolutions are concatenated, in the natural order of resolutions ($2,3,\ldots,B-1$ bits), along with a final network for the target resolution $B$, to produce the desired $B$-bit resolution solutions to problem~\eqref{opt_2}.

The proposed PCNet is trained in an unsupervised manner with a loss function derived from \eqref{class_obj} but based on the pre-softmax ${\mathbf B}_{{\mathbf f}_{{\rm RF}_k}}$ and ${\mathbf B}_{{\mathbf w}_{{\rm RF}_k}}$ instead of the binary ${\mathbf A}_{{\mathbf f}_{{\rm RF}_k}}$ and ${\mathbf A}_{{\mathbf w}_{{\rm RF}_k}}$. For the $B=2$ network, the loss function is
\begin{equation}
\label{eq:loss2bit}
\mathcal{L}_{B=2} (\Theta;\mathbf{H}_k) = -\left|({\mathbf B}_{{\mathbf w}_{{\rm RF}_k}}{\mathbf p}_{{\mathbf w}_{{\rm RF}_k}})^{\it{H}} \mathbf{H}_k ({\mathbf B}_{{\mathbf f}_{{\rm RF}_k}}{\mathbf p}_{{\mathbf f}_{{\rm RF}_k}})\right|\Big|_{B=2} 
\end{equation}
where $\Theta$ represents all trainable parameters in the proposed PCNet. For the $B=3$ network, the loss function is
\begin{align}
\label{eq:loss3bit}
\mathcal{L}_{B=3} & (\Theta;\mathbf{H}_k) \nonumber\\
= &-\left|({\mathbf B}_{{\mathbf w}_{{\rm RF}_k}}{\mathbf p}_{{\mathbf w}_{{\rm RF}_k}})^{\it{H}} \mathbf{H}_k ({\mathbf B}_{{\mathbf f}_{{\rm RF}_k}}{\mathbf p}_{{\mathbf f}_{{\rm RF}_k}})\right|\Big|_{B=2} \nonumber\\ 
&-\left|({\mathbf B}_{{\mathbf w}_{{\rm RF}_k}}{\mathbf p}_{{\mathbf w}_{{\rm RF}_k}})^{\it{H}} \mathbf{H}_k ({\mathbf B}_{{\mathbf f}_{{\rm RF}_k}}{\mathbf p}_{{\mathbf f}_{{\rm RF}_k}})\right|\Big|_{B=3}
\end{align}
which is used to update the parameters of the entire concatenated network. 

\section{Simulation Results}
\label{sec:results}
\subsection{Simulation Settings}

We simulate a multiuser mmWave massive MIMO system with $N_t=64$, $N_t^{\rm RF}=8$, $N_r=16$, $N_r^{\rm RF}=1$, and $K=8$. We consider the channel model with $L_k=10$ propagation paths for each MS. The azimuth and elevation angles of arrival and departure of each propagation path are assumed to follow the Laplacian distribution with uniformly distributed mean angles over $[0,2\pi]$ and angular spread of $10$ degrees. The signal-to-noise ratio (SNR) is defined as ${\rm SNR}= \frac{P}{K\sigma^2}$, where $P$ is set to $1$. The proposed PCNet is compared with the following benchmarks: 
\begin{itemize}
\item {\it FullDigital:} the traditional fully-digital beamforming scheme; 
\item {\it LowComplexity \cite{low}:} the principal component analysis (PCA)-based hybrid beamforming algorithm for designing analog precoder and combiner for specifically 2-bit resolution PS, plus the MMSE baseband precoder to handle the multiuser interference;
\item {\it SVD \cite{SVD}:} the singular value decomposition (SVD)-based hybrid beamforming algorithm for first designing the analog combiner and then the analog precoder, plus the zero-forcing (ZF) baseband precoder to manage the multiuser interference;
\item {\it JointDesign \cite{JD}:} the channel-decomposition-based hybrid beamforming algorithm for first designing the analog precoder and then the analog combiner, plus the ZF baseband precoder;
\item {\it CrossEntropy \cite{CEO}:} the cross-entropy-based hybrid beamforming algorithm for designing the analog precoder and combiner with finite-resolution PSs in an iterative manner, plus the ZF baseband precoder.
\end{itemize}
Note that the LowComplexity scheme was originally proposed for $B=2$ only and thus is compared in this setting only. SVD and JointDesign are originally infinite-resolution algorithms and are adapted to finite-resolution settings by quantizing their infinite-resolution solutions to the nearest point in the discrete phase set. The numbers of iterations, candidates, and the smoothing parameter for CrossEntropy are set to $20$, $150$, and $0.8$ when $B=2$, and $30$, $150$, and $0.8$ when $B=3$, respectively. To train the PCNet, an Adam optimizer is employed. The initial learning rate is set to 0.00003 and the batch size is set to 256. The numbers of channel samples in the training dataset, validation dataset, and testing dataset are set to $180000$, $20000$, and $10000$, respectively. Note that after the training process, the trained weights were recorded and re-training was not required for all possible channel states.

%%%%%%%%%%%%%%%%%%%%%%%%%%%%%%%%%
\begin{figure}[!h]
\centering
\includegraphics[width=0.9\columnwidth]{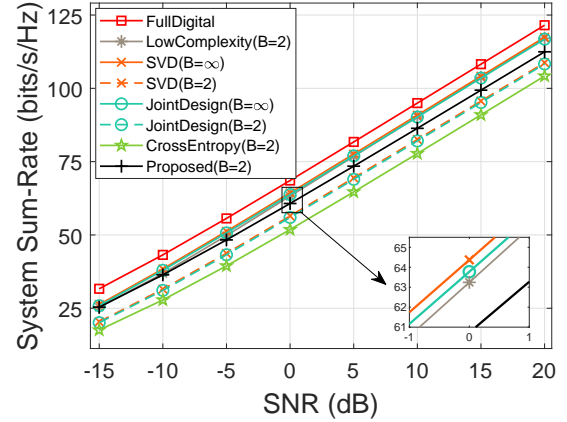}
\caption{System sum-rate vs. SNR for $B=2$. }
\label{fig:SNR2bit}
\end{figure}
%%%%%%%%%%%%%%%%%%%%%%%%%%%%%%%%%%
\begin{figure}[!h] 
\centering
\includegraphics[width=0.9\columnwidth]{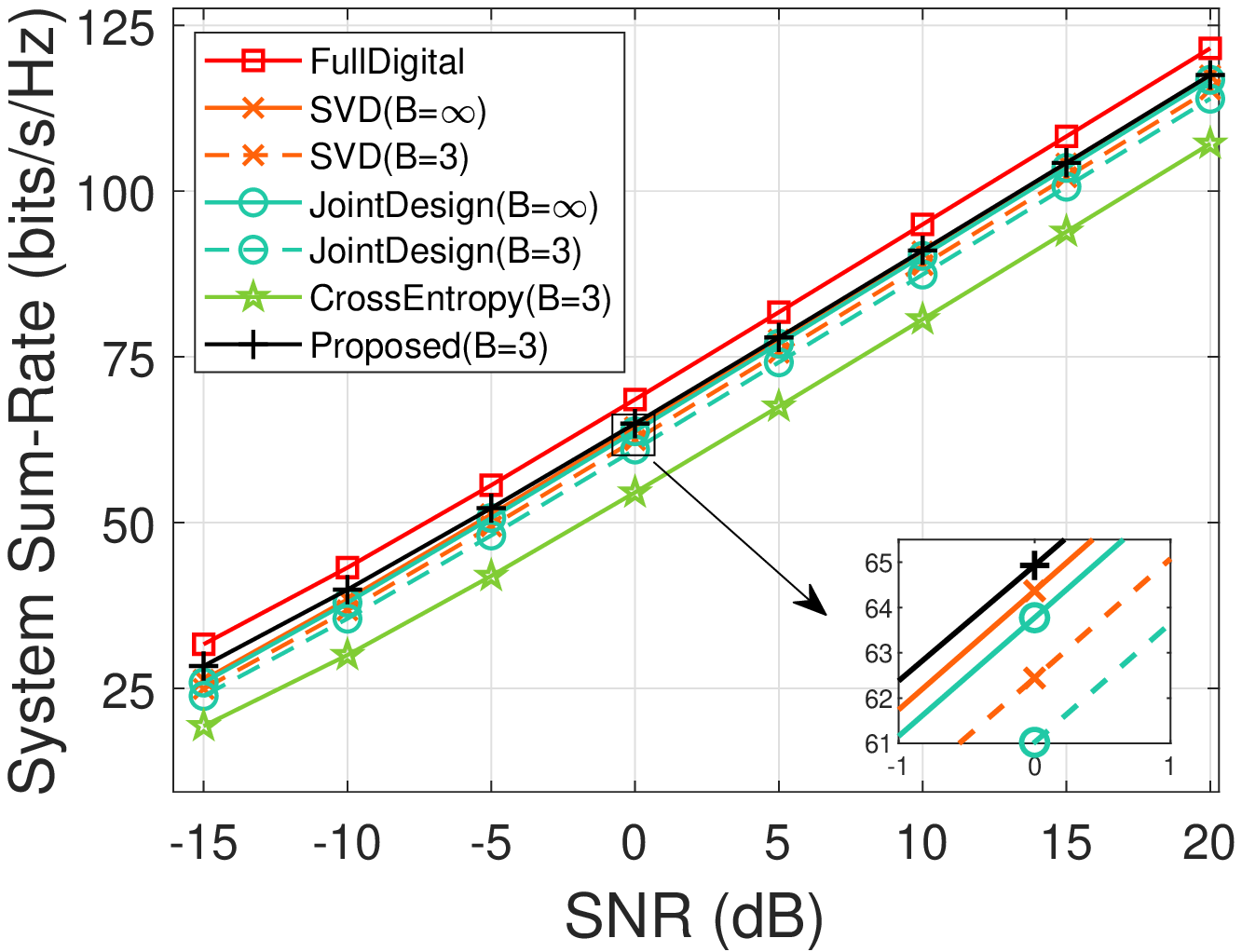}
\caption{System sum-rate vs. SNR for $B=3$.}
\label{fig:SNR3bit}
\end{figure}
%%%%%%%%%%%%%%%%%%%%%%%%%%%%%%%%%%%%
\begin{figure}[t] 
\centering
\includegraphics[width=0.9\columnwidth]{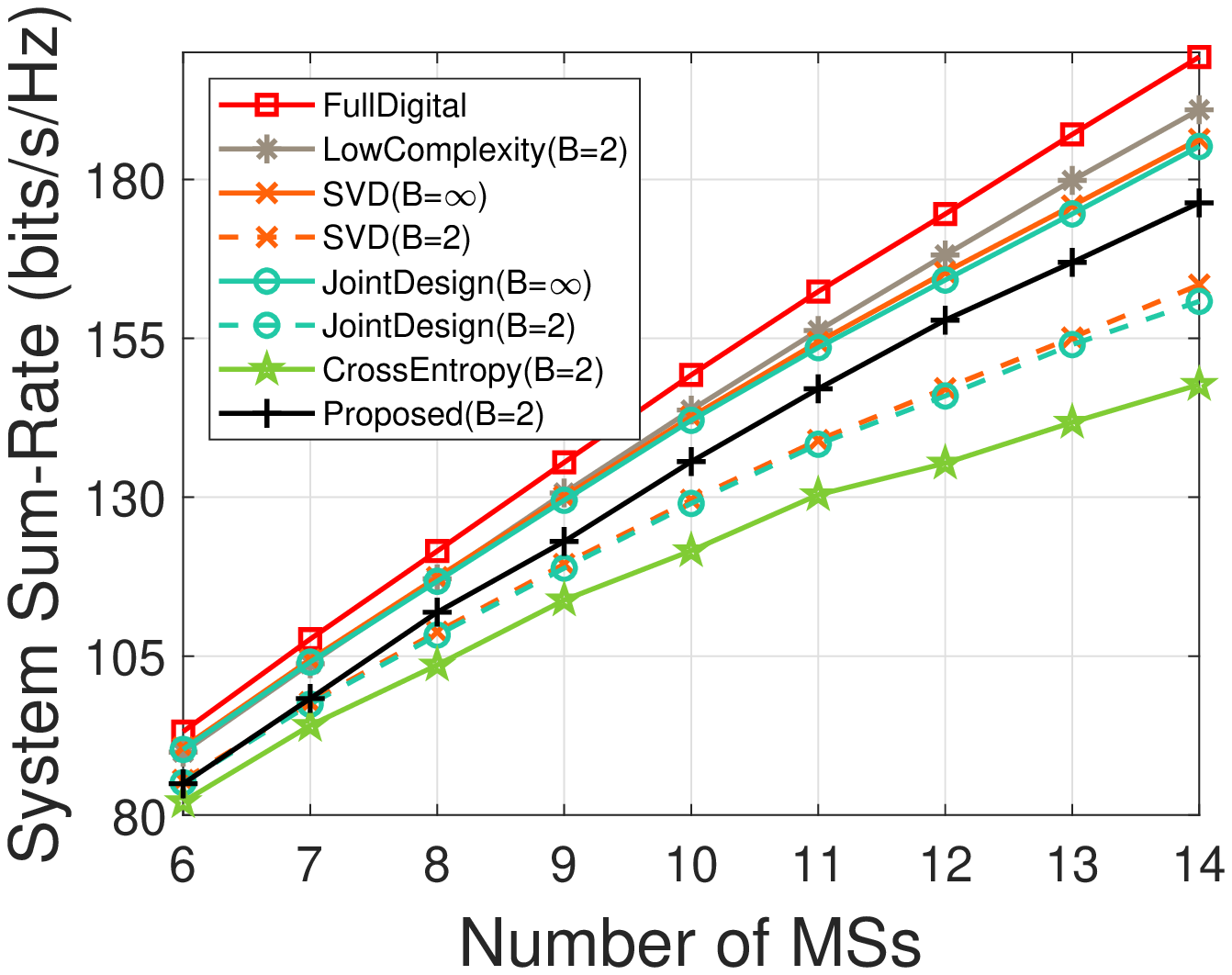}
\caption{System sum-rate vs. number of MSs for $B=2$ at $\mbox{SNR}=20$ dB.}
\label{fig:UE2bit}
\end{figure}
%%%%%%%%%%%%%%%%%%%%%%%%%%%%%%%%%%%%
\begin{figure}[h] 
\centering
\includegraphics[width=0.9\columnwidth]{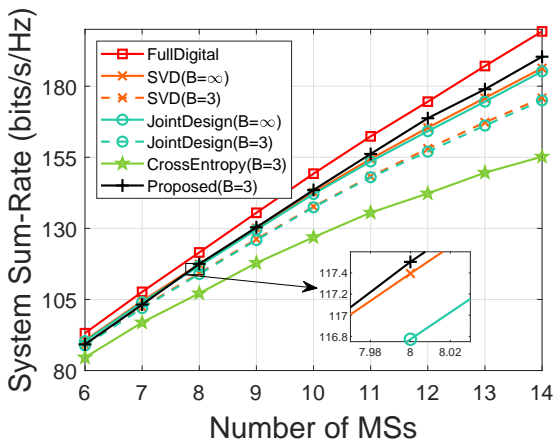}
\caption{System sum-rate vs. number of MSs for $B=3$ at $\mbox{SNR}=20$ dB.}
\label{fig:UE3bit}
\end{figure}
%%%%%%%%%%%%%%%%%%%%%%%%%%%%%%%%%%%%

\subsection{Results and Discussion}

Fig.~\ref{fig:SNR2bit} and Fig.~\ref{fig:SNR3bit} plot the system sum-rate vs. SNR performance for $B=2$ and $B=3$, respectively. The algorithms originally proposed for $B=\infty$ are also shown with the $B=\infty$ configuration for comparison (i.e., SVD and JointDesign). FullDigital serves as a performance limit for all schemes. As seen in Fig.~\ref{fig:SNR2bit}, the proposed PCNet outperforms SVD ($B=2$) and JointDesign ($B=2$), since the proposed PCNet directly designs finite resolution PSs while others suffer from quantization degradation. The proposed PCNet also outperforms CrossEntropy ($B=2$), because CrossEntropy adopts an iterative mechanism and could get stuck in a local optimum. The proposed PCNet is outperformed by LowComplexity ($B=2$) which was designed for specifically $B=2$ PSs and not easily generalizable. The proposed PCNet is outperformed by SVD ($B=\infty$) and JointDesign ($B=\infty$) because naturally infinite resolution PSs provide more degrees of freedom for beamforming and thus better performance.

In Fig.~\ref{fig:SNR3bit}, the proposed PCNet for $B=3$ outperforms SVD ($B=\infty$) and JointDesign ($B=\infty$). This may be attributed to the proposed concatenated design that facilitates exploiting the $B=2$ result to obtain a better $B=3$ result, and that adopts a loss function incorporating $B=2$ and $B=3$ contributions to train the end-to-end network for $B=3$. The proposed PCNet achieves the best performance among all $B=3$ schemes.

In Fig.~\ref{fig:UE2bit}, the system sum-rate vs. the number of MSs performance is shown for $B=2$ and $\mbox{SNR}=20$ dB. As can be seen, the system performance of all algorithms improves as $K$ increases. However, the gap between FullDigital and SVD ($B=2$), JointDesign ($B=2$), and CrossEntropy ($B=2$) increases as $K$ increases. This is because these algorithms adopt ZF to manage multiuser interference and thus suffer from performance degradation as the number of users increases. In contrast, the proposed PCNet and LowComplexity employ the MMSE precoder, and thus hold consistent gap with respect to FullDigital. Moreover, the proposed PCNet outperforms SVD ($B=2$), JointDesign ($B=2$), and CrossEntropy ($B=2$) algorithms over different numbers of users, confirming the robustness of the proposed method for different numbers of users.
%especially with high MSs, the performances of latter two tend to saturate. On the contrary, the performances of the proposed PCNet grow linearly with $K$.

Fig.~\ref{fig:UE3bit} illustrates the performance with the same setting as Fig.~\ref{fig:UE2bit} but for $B=3$. The proposed PCNet exceeds the performance of SVD ($B=\infty$) and JointDesign ($B=\infty$). As previously mentioned, the proposed PCNet achieves high performance for $B=3$ due to the concatenated model design of PCNet, enabling PCNet to effectively utilize previously acquired information. The superiority of PCNet holds with different numbers of users. 

Finally, Table~\ref{tab:Time} lists the average execution time for all algorithms, which does not include the training time of neural networks. For $B=2$, the proposed PCNet trails LowComplexity by a small margin in the sum-rate performance, but has over $20$ times lower complexity. The proposed PCNet achieves better performance {\it and} lower complexity as compared to other schemes. For $B=3$, the proposed PCNet achieves the best performance {\it and} lowest complexity among all schemes. When $B$ increases from $2$ to $3$, the complexity of all schemes increases, but by different amounts. Specifically, the complexity of CrossEntropy nearly doubles since the number of iterations therein increases from $20$ to $30$. The complexity of SVD and JointDesign only increases slightly since the execution time is dominated by the beamforming process instead of the quantization process. The complexity of the proposed PCNet increases only by $40\%$ as the complexity is dominated by the MMSE precoder instead of the increased NN architecture size. Hence, the concatenated network design incorporating lower-resolution networks provides superior performance without introducing heavily increasing complexity. Note that the proposed PCNet exhibits attractive performance-complexity tradeoffs as compared to $B=\infty$ schemes, and is more practical.

%%%%%%%%%%%%%%%%%%%%%%%%%%%%%%%%%%%%%
\begin{table}[!t] 
\caption{The Average Execution Time (N/A Means ``Not Applicable'')}
\label{tab:Time}
\centering
\begin{tabular}{|l|ccc|}
\hline
\diagbox[innerleftsep=.5cm]{Scheme}{B} & $\infty$ & 2 & 3  \\
\hline
\text{FullDigital} & 12.33 ms & N/A & N/A  \\
\hline 
\text{LowComplexity\cite{low}} & N/A & 25.07 ms & N/A  \\
\hline 
\text{SVD\cite{SVD}} & 1.64 ms & 2.8 ms & 2.97 ms \\
\hline 
\text{JointDesign\cite{JD}} & 2.23 ms & 2.94 ms & 3.01 ms  \\
\hline 
\text{CrossEntropy\cite{CEO}} & N/A & 37.1 ms & 52.9 ms  \\
\hline 
\text{Proposed PCNet} & N/A & 1.07 ms & 1.48 ms  \\
\hline
\end{tabular}
\end{table}
%%%%%%%%%%%%%%%%%%%%%%%%%%%%%%%%%%%%%%%%%%%%%%%%%%%%%%%%%%%%%%%%%

\section{Conclusion} \label{sec:conclusion}

We have proposed an unsupervised learning-based hybrid beamforming algorithm for MU-MIMO systems. The proposed algorithm incorporates a concatenated neural network design for low-resolution PSs, where lower-resolution solutions are exploited to produce better higher-resolution solutions. The scheme is applicable to designing PSs with arbitrary resolutions. Simulation results demonstrated that the proposed scheme can approach and even exceed the performance of infinite-resolution algorithms, with significantly lower complexity, and is superior over state-of-the-art finite-resolution hybrid beamforming designs in terms of performance-complexity tradeoffs.

\bibliographystyle{IEEEtran}
\bibliography{IEEEabrv,ref}
\balance

\end{document}